\documentclass[conference]{IEEEtran}
\usepackage[cmex10]{amsmath}
\usepackage{amssymb}
\usepackage{mathtools}
\usepackage{amsfonts}
\usepackage{listings}
\usepackage{color}
\definecolor{mygreen}{rgb}{0,0.6,0}
\definecolor{mygray}{rgb}{0.5,0.5,0.5}
\definecolor{mymauve}{rgb}{0.58,0,0.82}

\usepackage{multirow}
\usepackage{array}
\ifCLASSINFOpdf
   \usepackage[pdftex]{graphicx}
\else
\fi
\begin{document}
%
\title{$k$-rAC -- a Fine-Grained $k$-Resilient Access Control Scheme for Distributed Hash Tables}

\author{\IEEEauthorblockN{Olga Kieselmann}
\IEEEauthorblockA{Applied Information Security, University of Kassel\\
olga.kieselmann@uni-kassel.de}
\and
\IEEEauthorblockN{Arno Wacker}
\IEEEauthorblockA{Applied Information Security, University of Kassel\\
arno.wacker@uni-kassel.de}
}

\maketitle

\begin{abstract}
Distributed Hash Tables (DHT) are a common architecture for decentralized applications and, therefore, would be suited for privacy-aware applications.
However, currently existing DHTs allow every peer to access any index.
To build privacy-aware applications, we need to control this access.
In this paper, we present $k$-rAC, a privacy-aware fine-grained AC for DHTs.
For authentication, we present three different mechanisms based on public-key cryptography,  zero-knowledge-proofs, and cryptographic hashes.
For authorization, we use distributed AC lists. The security of our approach is based on $k$-resilience.
We show that our approach introduces an acceptable overhead and discuss its suitability for different scenarios.
\end{abstract}


%

\section{Introduction}
\label{sec:introduction}
A Distributed Hash Table (DHT) is a convenient way to utilize the power of highly scalable Peer-to-Peer (P2P) networks. Currently, there exist numerous ways to build such a DHT, e.g., Chord~\cite{stoica2001chord}, CAN~\cite{ratnasamy2001scalable}, Kademlia~\cite{maymounkov2002kademlia}. Some of those are used in existing products, mostly for file sharing. For instance, BitTorrent~\cite{loewenstern2008bep} uses Kademlia to find other peers offering the same file (trackers). In general, the scalability and the architecture of DHTs are well researched. However, security issues like access control or storing confidential data in such a DHT are still mostly unsolved. While the existing solutions are suitable for currently existing products, there are scenarios where we require a reliable and lightweight access control for protecting confidential data. Access control is usually defined as the combination of authentication and authorization~\cite{stamp2011information}. The purpose of authentication is to determine whether a user is allowed to access a system. In contrast, authorization regulates the access to various system resources.

Applying this general definition to an access control for a DHT, authentication means the regulation of the participation in the P2P network. In the literature, this is referred to as a coarse-grained access control. Accordingly, authorization regulates the access to entries in the DHT. In our work, we additionally separate user permissions for reading or writing each single entry. By combining the authentication with this expanded authorization, we consider the access control as fine-grained.

However, implementing such a fine-grained access control for a DHT is challenging due to the architecture of a DHT. Generally, DHTs offer the following two operations:

\begin{itemize}
  \item put (key, value) - store a key-value pair for a certain key, i.e., write a value,
  \item get (key) - retrieve the value for a certain key, i.e., read a value.
\end{itemize}

Without further modification, any peer can read or write any key via the above operations. Additionally, peers have full access to entries under their own control. We must consider that some peers might act maliciously, e.g., manipulate values under their control. Therefore, we need to control the read and write access to each individual entry of the DHT.

A scenario where a fine-grained access control is required is the Social Link project \cite{david2014facilitating}. Their goal is a new communication paradigm for supporting the work-life balance based on appraisal of users' context for the implicit communication. A DHT is used for storing the users contexts with sensitive data, e.g., localization, activity or calendar data. Therefore, access to this data must be secured in such a way that it should be possible to assign the write and/or read access to specific users. Another scenario is the Data Revocation Service~(DRS)~\cite{kikowa2015drs} for supporting the deletion of personal data objects on the Internet. With the DRS, users store status information of their data objects in a DHT. By requesting this status for a certain data object with the DRS, providers verify whether they are allowed to deliver this data object. Since the status information must be accessible for anyone, only the write access must be protected for updating the status information by the authorized user. Another scenario are P2P-based massively multiuser virtual environments (MMVEs) allowing many users to participate in a shared virtual environment via the Internet. MMVEs can be implemented as a DHT as proposed in~\cite{wacker2008towards}. The participants of a large scale system typically do not know each other and, therefore, cannot trust each other, since it is possible that users may pose as somebody else or steal other users' data.

Although there exist approaches to implement an access control scheme for P2P networks, to the best of our knowledge, there is none combining the following features:
\begin{itemize}
    \item individual restriction for the read and write access to a single DHT entry,
    \item revocation of put or get access for a certain user,
    \item delegation of administration for put and get access,
    \item guaranteed resilience against up to $k$ malicious peers, where $k$ is a system parameter.
\end{itemize}


Hence, in this paper we present $k$-rAC, a fine-grained access control to read or write any key, even if a certain number of peers are subverted by an attacker. Additionally, $k$-rAC allows the delegation of access rights to other users. The contribution of this paper is twofold: First, we propose three different generic access control mechanisms for $k$-rAC to regulate the read and write accesses in a modular way (cf. Section~\ref{sec:ac}). Secondly, we provide detailed evaluation of these mechanisms regarding the additional effort in comparison to a classical DHT without any access control.


The rest of this paper is organized as follows: We first discuss existing access control mechanisms for P2P networks in Section~\ref{sec:relatedwork}. In Section~\ref{sec:systemmodel}, we present our assumptions, and motivate our requirements for a privacy-aware $k$-resilient access control in P2P networks in Section~\ref{sec:requirements}. Based on those, we discuss the design rationale of $k$-rAC in Section~\ref{sec:dr} and the novel access control scheme in Section~\ref{sec:ac}. Afterwards, in Section~\ref{sec:evaluation}, we analytically evaluate our approach and determine its performance by simulation. Finally, we conclude our paper in Section~\ref{sec:conclusion} with a brief summary and provide an outlook on future research.

\section{Related Work}
\label{sec:relatedwork}

As mentioned above, access control schemes are comprised of authentication and authorization. Authentication mechanisms have been already studied extensively and there are reliable approaches available. The basic idea of these approaches is the usage of a public key infrastructure (PKI). Hereby, each peer uses a signed certificate from a certification authority (CA) to authenticate itself. Furthermore, it can be used to realize confidential communication between arbitrary peers. This coarse-grained access control is used, e.g., in~\cite{waste}. However, this approach treats all peers equal and, thus, allows all authenticated peers to access any information in the system. In~\cite{tamassia2007efficient}, the authors also proposed an authentication-only approach but based on a distributed Merkle tree. They focused on achieving consistent updates in the DHT and resilience against replay attacks. The authors of~\cite{lu2008pseudo} pursue an anonyme authentication mechanism based on a Zero-Knowledge Proof~\cite{feige1988zero}. Takeda et al.~\cite{takeda2008new} propose a decentralized mutual authentication mechanism for each pair of nodes to avoid performance issues of a PKI approach. In~\cite{jawad2009protecting}, the authors build on the reputation-based trust management to realize the authentication of peers. Here, a DHT is used only for storing the trust levels of each user. The authorization part is solved by storing the data locally on the owners computer. The data owner requests the trust levels from the DHT to decide whether to allow the access to data on her computer for a certain user. In~\cite{berket2004pki}, the authors extend the PKI approach by trusted groups to regulate the access to resources based on group memberships. MacQuire et al.~\cite{macquire2008authentication} propose an improved routing protocol for DHTs tailored for highly heterogeneous peers. The main idea of their approach is that peers with different capabilities have different roles in the DHT. To do so, they also extend the PKI-approach to include permissions for authenticated peers. However, the above approaches do not consider the fine-grained protection of the get and put operations for individual DHT entries.

In \cite{kubiatowicz2000oceanstore}, the authors propose OceanStore, an architecture for a global-scale persistence storage. Regarding the authorization, they mention that the data should be encrypted. Although they propose to use access control lists (ACL) for the reader and writer restrictions, they do not propose a specific solution how to protect this ACL from manipulation. In general, they provide only abstract details about the realisation of their access control, as their focus is on the global architecture. Another difference is that they offer read protection only out-of-band, i.e., the users need to distribute encryption keys. In contrast, we specified the details of our access control and offer an in-band key exchange for the read protection.
The proposed fine-grained access control mechanism in~\cite{sturm2008access} is build on a hierarchical model of peers. Here, the developers connect servers with dozens of users into a P2P network and do not consider home office computers. A single server manages the access control policies of its users and collaborates with other servers to delegate access control to other peers and users. Therefore, the proposed approach is closely tied to their scenario and requires a central server. In contrast, our approach is a general access control for a DHT and fully decentralized. Furthermore, the proposed approaches in~\cite{isdal2010privacy, tang2010fade} are not applicable for regulating the fine-grained access control in a DHT, since they also use centralized components to manage access policies.

There are approaches with similar goals as ours. In~\cite{palomar2006certificate}, Palomar et al.\ propose a PKI-based access control scheme by extending certificates with authorization capabilities. However, their approach relies on trusted groups and does not allow for a fine-grained access control where individual permissions can be set for each entry in the DHT. In P-Hera~\cite{crispo2005p}, the access control scheme allows data owners to specify fine-grained restrictions on who can access their data. For this, they use super nodes to manage access policies. In contrast, our approach does not rely on any centralized components. In~\cite{narendula2009towards}, they use ACLs for controlling the access to individual keys of a DHT. However, they protect these ACLs by using trusted groups, whereas we propose to use the more general $k$-resilience. Further, in~\cite{sanchez2010distributed}, the authors propose a protocol for delegating access control to intermediaries in such a way that requesters do not learn the access policy, and the intermediates do not learn the privileges. Although the authors' focus is privacy, they do not distinguish between put and get operation to secure the access to a single DHT entry.

The closest work to $k$-rAC is DECENT~\cite{6197504} that proposes an architecture for enforcing access control in a decentralized online social network (OSN).
Although based on a DHT, this architecture is specifically tailored to the OSNs. Similar to us, the authors aim to regulate the read and write access to a single DHT entry, and to delegate access rights to other users. To achieve that, they use public key cryptography and describe how to apply it for the OSN scenario. We, in contrast, propose three different access control mechanisms: one is also based on the public key cryptography, the second is based on the zero knowledge proof, and the third one is derived from password hashes. The three mechanisms are all generic for the usage in a DHT for arbitrary scenarios. Moreover, we compare them with each other and consider in which scenarios they suit better.
To cope with malicious peers, the authors rely on replication. However, they do not mention any further details or specify a protection method against malicious peers.  With $k$-rAC, we propose a specific mechanism to enable resilience up to $k$ malicious peers. Hence, with $k$, we introduce a parameter to achieve a certain security guarantee. Furthermore, we evaluate our access control scheme by detailing the effort for the three access control mechanisms. With our evaluation, researcher and engineers are able to decide which particular mechanisms is suitable for a specific scenario. Contrarily, the authors of DECENT evaluated only the public key approach for the OSN scenario. 

In summary, the existing approaches are either limited to a certain scenario or do not offer a viable solution for fine-grained access control. In contrast, we propose a fine-grained access control for DHTs without the need of any centralized components, trusted groups, or super nodes.

\section{System Model}
\label{sec:systemmodel}
Our system is a P2P-based network forming a DHT based on any of the existing approaches. This DHT must cope with churn, scalability, availability, persistence, consistence and routing, e.g., Kademlia~\cite{maymounkov2002kademlia}. With other words, we do not propose a new DHT but an extension to an arbitrary existing one. Participating \emph{peers} offer the operations put and get (cf.\ Section~\ref{sec:introduction}) to the application, i.e., to \emph{users}. Peers implement the DHT-algorithm and interact with each other via messages over an arbitrary network, e.g., the Internet. In a DHT, each peer has a unique ID determining its position in the DHT space. We assume that the ID is determined by the system, and the peer cannot influence it. This can be achieved by, e.g., hashing the IP address. Further, there is an authentication mechanism for the peers, e.g., a CA issuing individual certificates to each peer as proposed in~\cite{narendula2009towards}. Every message sent from a peer is signed with its own individual certificate. Thus, we can verify the authenticity and the integrity of each received message. Any message without a valid signature is discarded. By doing so, only the authenticated peers can participate in the network. Additionally, confidentiality can be achieved.

\begin{figure}
  \centering
  \includegraphics[width=0.8\columnwidth]{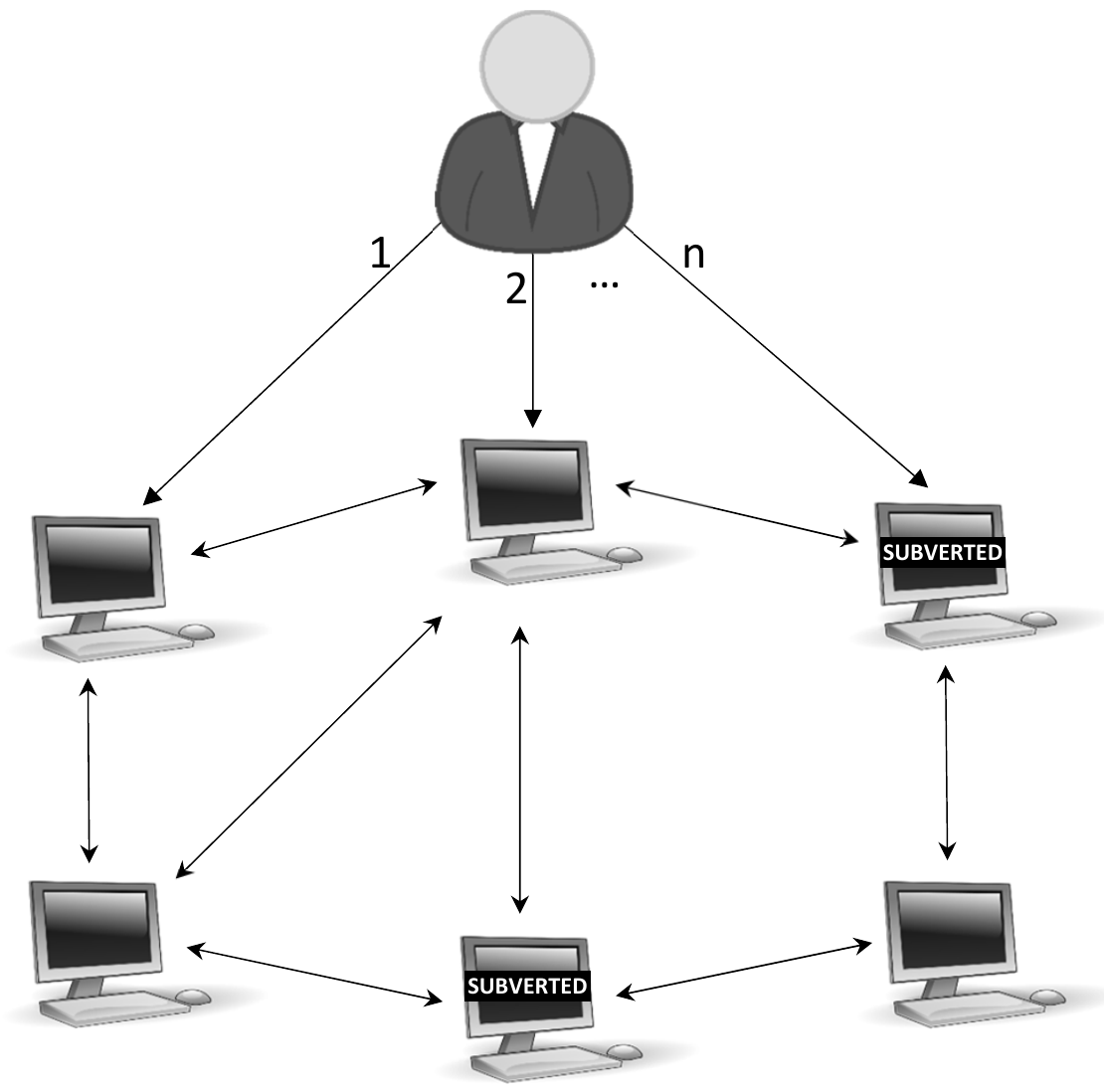}\\
  \caption{System Model}\label{fig:systemmodel.user}
\end{figure}

The DHT space comprises a certain number of \emph{DHT entries}. A DHT entry consists of an \emph{index} and a \emph{value}. The index, also referred to as the key, uniquely identifies an entry. However, for the rest of this paper, we use the term \emph{index} to avoid the confusion with cryptographic keys. For each index in the DHT, there is at least one \emph{responsible peer} for storing the DHT value. The value can contain additional metadata about the stored object. To differentiate, we call the entire object stored under a certain index a \emph{DHT value} or \emph{value}. When we refer only to the stored data, we call this the \emph{data object}.

The user interacts with the DHT via the application interface (API) of one or more arbitrary peers. In our system, we assume that the application software (i.e., the user) does not necessarily run on the same node as the P2P-software (i.e., the peer). The application software accesses the peers via secure channels (e.g., SSL/TLS). Hence, we assume a clear separation of users and peers (cf.\ Figure~\ref{fig:systemmodel.user}). Specifically, the user calls the put or get operation in order to store or retrieve a value, respectively. Furthermore, we assume that our application requires a fine-grained access control. A user is allowed to read or write only those DHT entries to which she has the appropriate rights.

We assume the presence of an \emph{attacker} with the goal to circumvent the access control, i.e., to gain access to values in the DHT without the proper access rights. For this, the attacker is able to compromise arbitrary peers of the underlying P2P network. For an external attacker, we rely on existing solutions to regulate the access to the network with a CA (e.g.,~\cite{berket2004pki}). Furthermore, we assume that the attacker can subvert at most $k$ arbitrary peers at any time. A subverted peer is under full control of the attacker, specifically, the attacker has access to the individual certificate of the subverted peer. Therefore, the attacker can send, receive, discard, replay, and forge messages on behalf of the peer. However, we assume that a user is not compromised, i.e., the attacker cannot subvert the application software.

\section{Requirements}
\label{sec:requirements}
In this section, we describe and motivate the requirements for our novel access control scheme.

\textbf{Access (A):} A user must be able to perform only the actions if and only if she has the appropriate rights.

\textbf{Ownership (O):} For our fine-grained access control scheme, we require a mechanism to uniquely determine the owner of a certain DHT entry. Once the ownership is set, it must be impossible for an attacker to steal the ownership. Only the owner must be able to revoke her ownership.

\textbf{Granularity (G):} The owner must be able to manage read and write right separately for each of her entries in the DHT. These rights can be assigned to individual users. Additionally, the owner must also be able to delegate the handling of read and write rights to other users.

\textbf{Privacy (P):} When protection of user's privacy and anonymity is required (e.g.,~\cite{kikowa2015drs}), it must be impossible to deduce the owner from the DHT entries. Specifically, our scheme must not introduce \emph{new} ways to deduce the owner of a DHT entry. Further, our scheme must not provide a way for any entity to find all DHT entries of a certain owner. In general, our scheme must not introduce \emph{new} privacy risks for anyone.

\textbf{Scalability (S):} DHTs are built with a high scalability in mind, i.e., they can handle a multitude of simultaneous read and write operations (get/put). Any access control scheme added on top of such a DHT must not degrade this property. Thus, the communication and processing overhead for any operation in the DHT must not increase significantly.

\textbf{$k$-Resilience (R):} In our P2P network, the attacker can compromise at most $k$ arbitrary peers. Therefore, we require that our access control scheme remains functional even if the attacker controls up to $k$ peers. Specifically, even in this worst case, our other requirements, i.e., A, O, G, P, and S, must still be fulfilled.

\section{Design Rationale}
\label{sec:dr}

Before enforcing a fine-grained access control as introduced in Section~\ref{sec:introduction}, we must determine the ownership of an entry in the DHT. This is necessary because by default each entry can be written and read by any peer (cf.\ Section \ref{sec:systemmodel}). To determine the ownership, we propose to use the approach from \cite{wacker2008towards} and determine the owner of an entry by the first access to this entry. Specifically, the user who first writes a value under a certain key becomes automatically the owner of this entry. After that, only she can access this value or delegate the access rights to other users.

Any operation in a DHT, i.e., put or get, is initiated by a user through a requesting  peer, routed through the network, and finally replied by a responsible peer. Hence, we can either control the access on the requesting peer or control the reply on the responsible peer. We follow the notation from ~\cite{narendula2009towards} and call this controlled queries (CQ) and controlled replies (CR). CQ are based on encrypted DHT values, and the access policy is enforced by providing the encryption key to authorized users. With CR, the responsible peer must only deliver the values according to the access policy. In $k$-rAC, to fulfill the $R$-requirement, we apply both concepts, i.e., CQ and CR. The rationale behind this is that the read access can only be achieved with CQ, while the write access can only be achieved with CR. In the following, we elaborate on this rationale.

With our $R$-requirement, controlling the read access cannot be achieved with CR since we must cope with the attacker compromising up to $k$ peers. By controlling only the replies, there might be a peer under the attacker's control which ignores the access policy and delivers the result anyway. Thus, the attacker would gain read access to the DHT entry, which violates our $A$-requirement. Therefore, we must enforce the read access with CQ. For this, the user encrypts her data before saving it in the DHT. After that, she distributes the encryption key to other users which are allowed to read the corresponding data. By doing so, the responsible peer does not have to verify whether the requesting user is authorized to read the data -- without the encryption key, she cannot read the encrypted data anyway. Here, the challenge is the distribution of the encryption key to other users. In Section~\ref{sec:ac}, we suggest how this challenge can be solved.

Similarly, we cannot enforce the write access with CQ, because the peer which intends to write might be under the attacker's control. A malicious peer could ignore the permissions and write the data in the DHT anyway, thereby violating our $A$-requirement. Therefore, we enforce the write access control on the receiver side, i.e., with CR.
To do so, we first authenticate the user and then verify whether she is authorized to write to the requested DHT index. The user authentication mechanism must also fulfil the requirements from Section~\ref{sec:requirements}. In Section~\ref{sec:ac}, we present several approaches how to realize the write access control.

Due to our $R$-requirement, the user never trusts a single peer, more specifically, she does not trust any group of up to $k$ peers. Therefore, to cope with up to $k$ subverted peers, she always interacts with the API of $2k+1$ different peers. Using these peers, she stores each DHT value at $2k+1$ different indexes. With evenly distributed indexes and sufficient peers participating in our P2P network, we assume that these $2k+1$ indexes are managed by different peers. All subsequent operations must be always executed on these $2k+1$ peers. Accordingly, while storing, each of these different responsible peers performs the access control. However, per definition at most $k$ do not. Hence, when performing a get operation, i.e., reading all $2k+1$ indexes, she receives at least $k+1$ correct values. Afterwards, she calculates a majority function over all received values and considers only the value which she received more than $k$-times.

\section{Access Control Scheme}
\label{sec:ac}

We propose to use an \emph{access control list} (ACL) for each DHT entry. For this, we assign an ACL to each non empty entry in the local storage of each responsible peer. A single ACL item contains the user's authentication \emph{auth} and her access rights. We define the following user access rights:

\begin{itemize}
    \item read (r) -- the right to read the value,
    \item write (w) -- the right to write the value,
    \item admin (a) -- the right to change, add, or remove the read or write access of other users,
    \item owner (o) -- this right implies r, w, and a. Additionally, it allows to set the a-right, i.e., the owner can pass the admin-right to other users.
\end{itemize}

While the owner is set by the system on first access, the other rights can be arbitrary managed by the owner or users with the admin right for a specific entry. This implies the access hierarchy $o > a > (r|w)$, where $>$ is defined as `includes right'. To verify these access rights, the user must first be authenticated. For this, we extend the standard DHT operations \emph{put} and \emph{get} with an additional parameter for the user authentication, named \emph{auth}. When a user performs an operation on the DHT (via the peer API), the responsible peer uses the user's \emph{auth} to verify whether she is authorized to execute the requested operation. Additionally, we introduce the new operation \emph{set} for solely managing the access rights on a DHT value. The put and the set operation are closely related as we could manage the access rights also with the put operation by including an \emph{acl} parameter. However, for the sake of simplicity, we preferred a clear separation of managing the data object and the ACL.
Hence, the operations for our access controlled DHT are:
\begin{itemize}
    \item put (index, data, auth) -- stores the data object \emph{data} at the \emph{index} in the DHT. The user is authenticated with the \emph{auth} parameter.
    \item get (index, $[$auth$]$) -- retrieves a data object from the given \emph{index}. The user authentication \emph{auth} is optional, since the read access control is done with CQ, i.e., it is enforced by distributing the encryption key.
    \item set (index, acl, auth) -- modifies the ACL without modifying the data object.
\end{itemize}

To implement this, we extend the DHT value to house the users' access rights besides the data. An empty entry has no ACL and is not owned by anyone. The first access via a put or set operation to a previously empty entry determines the owner of this entry. This means, the peer stores in the ACL the \emph{auth} of the user who first accessed this entry and set her access right to \emph{o}. Therefore, the smallest possible ACL is a list with the owner as a single item, whereby the stored \emph{auth} depends on the used authentication mechanism and will be described later. We integrate the ACL in the DHT value as presented in Listing~\ref{lst:acl_data_structure}.

\lstset{language=C,caption={Access Control Data Structures},label=lst:acl_data_structure}
\begin{lstlisting}[frame=single]
dht_value {
  acl: list<acl_item>;   // a list with acl items
  data: object;          // the stored data object
}

acl_item {
  auth: authenticator; // the user's authenticator
  key: encrypted_session_key; // data encryption key
  rights : {o,a,w,r};
}
\end{lstlisting}

As motivated in Section~\ref{sec:dr}, we propose that each DHT value is stored with $2k+1$ different peers and by using $2k+1$ different peer APIs. To achieve this, the user stores the value for the index \emph{index} within the DHT at positions $\mathit{pos}_i := h(\mathit{index}|i)$, for all $i$ with $1\leq i \leq (2k+1)$ and $i\in \mathbb{N}$. With the concatenation of \emph{index} and $i$, the user first calculates $2k+1$ different storing positions. However, these positions are consecutive and probably stored with the same peer. Hence, she additionally applies a cryptographic hash function $h(x)$ to evenly distribute all replicas of a value within the DHT space.

\subsection{User Authentication Mechanisms}
\label{sec:ac.authentication}
For the user authentication, we consider three different mechanisms with different properties. It depends on the specific scenario which of them suits better. Common to all three mechanism is their resilience against up to $k$ subverted peers. In each mechanism, the \emph{auth}-parameter in a DHT operation is used to authenticate the user.

\begin{figure*}
  \centering
  \includegraphics[width=350pt]{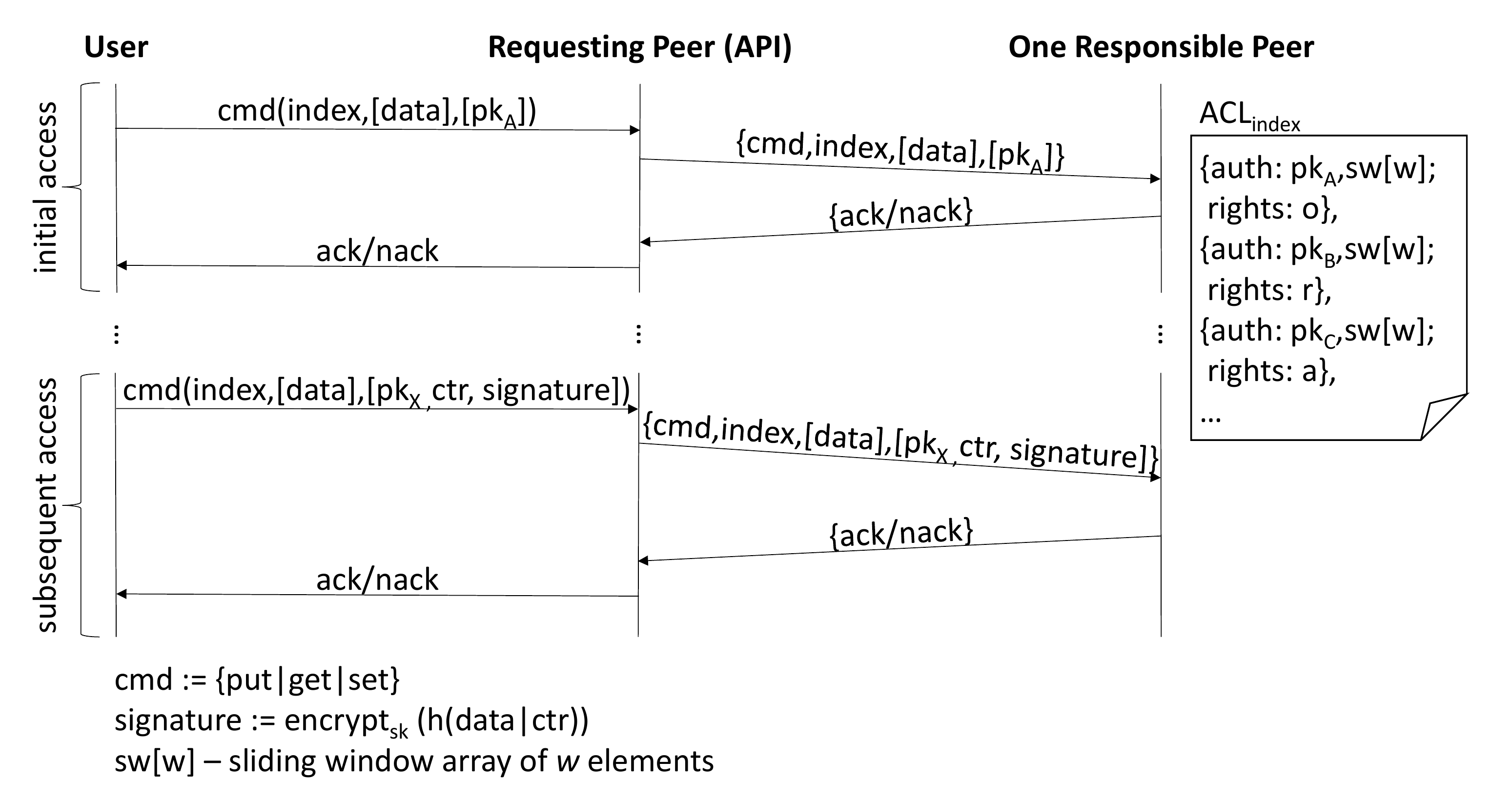}\\
  \caption{PK Message Flow}\label{fig:ac.authentication.PK}
\end{figure*}

\textbf{Public-key Cryptography}
Each user generates a public/private key pair $(pk, sk)$. According to that, the authenticator \emph{auth} includes the public key~$pk$, a counter $\mathit{ctr}$, and a signed hash of the data object concatenated with the counter, i.e., $\mathit{auth} := \{pk, \mathit{ctr}, \mathit{signature}\}$ where $\mathit{signature} := \textrm{encrypt}_{sk}(h(\mathit{data}|\mathit{ctr}))$ and $h(x)$ is an arbitrary cryptographic hash function. The responsible peers for the accessed index store upon first access the user's public key $pk$ in the \emph{auth} field of the ACL item. By doing so, this public key is now pinned to that index, and the user is defined as the owner (cf.~\ref{sec:ac.authorization}). In subsequent requests, the responsible peer verifies the validity of the signature with the stored public key. Only the user in possession of the right private key is able to generate a valid signature. To prevent replay attacks, we rely on the standard approach with a sliding window, similarly to IPsec~\cite{frankel2011ip}. With this approach, the peer stores at most $w$ old counters, where $w$ is the window size. The user also stores the counter locally. Thus, the freshness of the message is ensured, and the user is authenticated on subsequent requests for any operation with a valid signature. We summarize the message flow of the PK authentication in Figure~\ref{fig:ac.authentication.PK}.

Since the ACL does not contain personal information about the user, this approach does not leak information about user's identity. However, if a user owns multiple DHT indexes, an attacker could determine all her DHT entries by comparing the stored public keys. In some scenarios, this is not desirable and must be prevented, e.g.,~\cite{kikowa2015drs}. Here, the user should use different public/private key pairs for each index. This way, no connection between any two indexes can be established, even with an all-powerful attacker who could read the entire DHT.

\textbf{Zero-Knowledge Proof} With a zero-knowledge proof (ZKP)~\cite{goldwasser1985knowledge}, a prover proves to a verifier that she possess a secret without revealing it. It works as a challenge/response system, where the prover has a chance of $50\%$ to cheat in any single round. Therefore, the proof must be performed $n$ times to achieve a high confidence.

\begin{figure*}
  \centering
  \includegraphics[width=350pt]{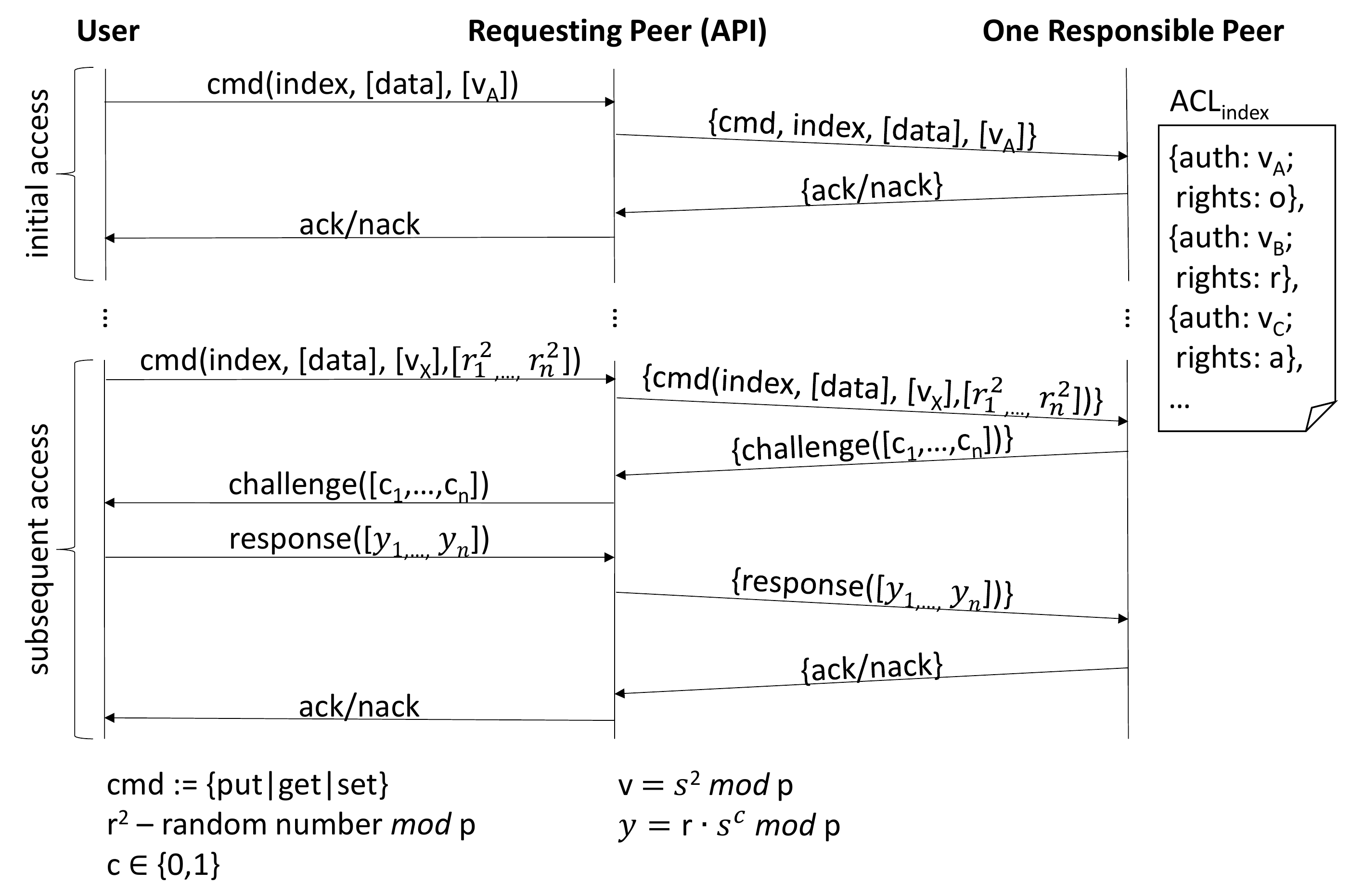}\\
  \caption{ZKP Message Flow}\label{fig:ac.authentication.ZKP}
\end{figure*}

We applied the ZKP to a DHT by building on the Feige-Fiat-Shamir protocol~\cite{feige1988zero}. Here, the user generates a secret random number $s$. The square of this secret, i.e., $v:= s^2\, \textrm{mod}\, p$, is stored as her authenticator on first access in the ACL item for the requested index, i.e., $\mathit{auth} := \{v\}$. On subsequent accesses, the received $v$ is used to identify the corresponding ACL item. If the received $v$ is not part of the ACL, the request is rejected. To authenticate, the responsible peer sends a message with $n$ challenges to the requesting user. If the user replies with $n$ correct responses, she is authenticated. Since each request is authenticated with new challenges/responses, the freshness of the messages is ensured and no additional mechanisms to prevent replay attacks are needed. After a successful authentication, the peer proceeds with the requested operation if this user has the appropriate rights (cf.~Section~\ref{sec:ac.authorization}). In Figure~\ref{fig:ac.authentication.ZKP}, we present the message flow of the ZKP authentication.

With this authentication scheme, the most costly part are the challenge/response messages for the authentication. As an optimisation, a peer could first check with the ACL whether the requesting user ist authorized to perform the corresponding operation for the requested index. If the user does not have the appropriate rights, the peer suspends the requested operation without performing the challenge/response.

Also with this approach, there is no personal information leaking about the user. The only information an attacker could get is the squared number $v$ used as the authenticator.
However, similarly to the public/private key approach, an attacker could determine all indexes belonging to the same user by comparing the $v$.
Again, this can be avoided by choosing different secrets for different DHT indexes.

\textbf{One-Time-Hash}
The classical way to authenticate a user in client/server systems is with password hashes. In such an authentication scheme, the user authenticates with a secret, i.e., her password. On the server side, only the hash of the password is stored. To verify whether the user has the correct password, she sends it to the server. Then the server hashes and compares it with the stored hash. If they match, the user is authenticated. To prevent the usage of precalculated hash tables, an additional salt~\cite{morris1979password} is usually stored. Salt is a random number, which is usually appended to the secret before hashing. To still verify the password, the \emph{salt} is stored in clear alongside the hash.

Due to our $R$-requirement, we cannot use this classical way to authenticate a user in a DHT. With this scheme, the user must submit her secret when calling our modified put, get, or set operations. The secret would be the user's authenticator for all $2k+1$ peers. By doing so, each peer which transports this message (at least all $2k+1$ responsible peers) would get to know the secret. Hence, even by subverting only one of these peers, the attacker would get access to the user's secret contradicting our $R$-requirement.

\begin{figure*}
  \centering
  \includegraphics[width=350 pt]{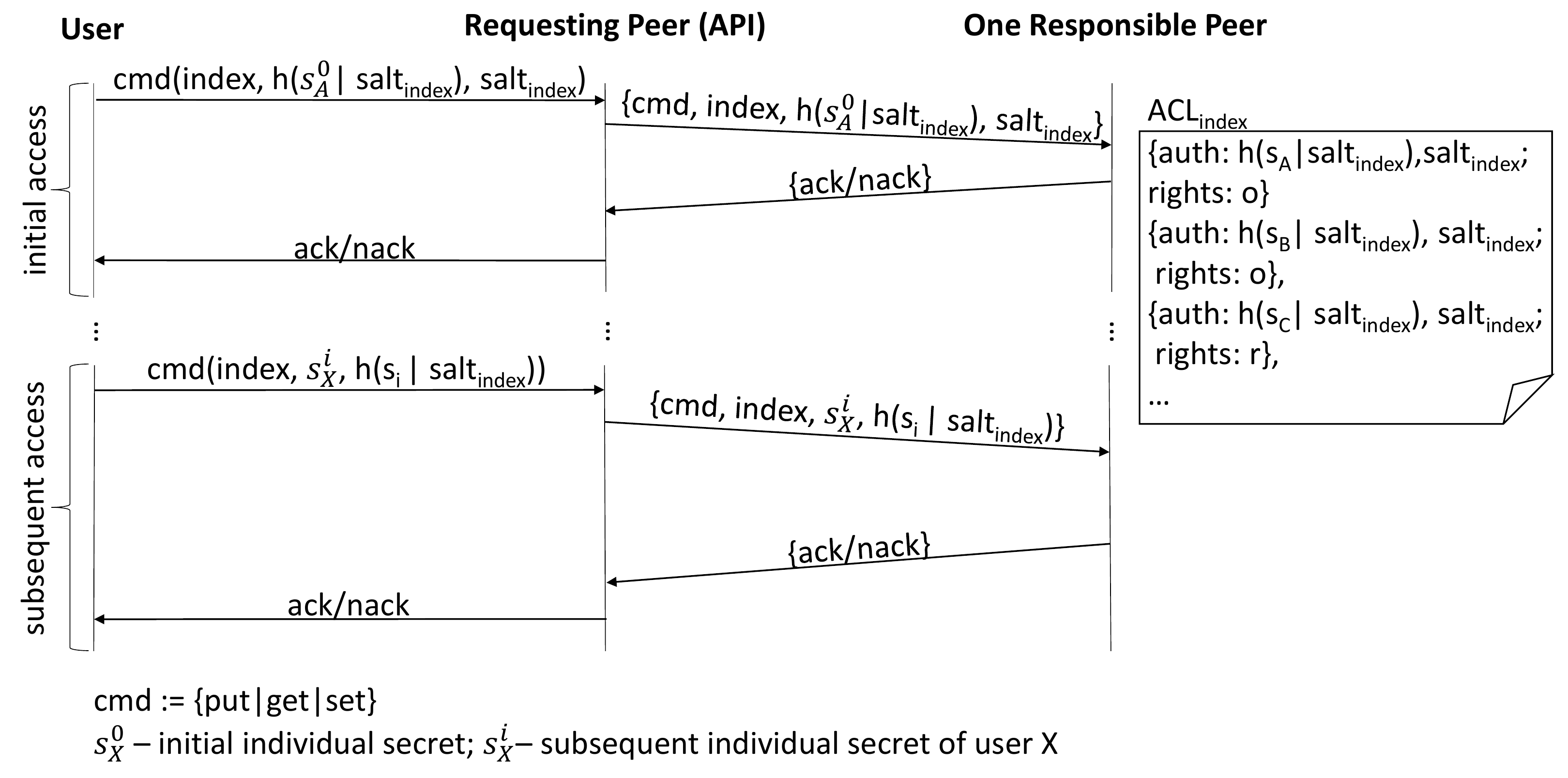}\\
  \caption{OTH Message Flow}\label{fig:ac.authentication.OTH}
\end{figure*}

To overcome this drawback, we extend the classical hash-based authentication by introducing an individual secret for each of the $2k+1$ peers, i.e., an individual authenticator for each of the responsible peers. However, instead of storing $h(s|\mathit{salt})$, the user uses an HMAC-function~\cite{krawczyk1997hmac} to create $2k+1$ individual secrets. Specifically, she calculates the individual secrets with $s_i = \textrm{HMAC}_s(\mathit{index}|i)$ for all $i$ with $1 \leq i \leq (2k+1)$, $i\in \mathbb{N}$ and sends the hashes of these secrets $h(s_i|\mathit{salt}_i)$ together with an individual $\mathit{salt}_\mathit{index}$ to the responsible peers.
She also stores the $\mathit{salt}_\mathit{index}$ for the $\mathit{index}$ locally.
For our $P$-requirement, we include the index in the calculation of these secrets resulting in different secrets for each index.
For the same reason, we use different salt-values for different indexes.
The responsible peers store the received hashes and $\mathit{salt}_\mathit{index}$ as the authenticator, i.e., $auth_i := \{h(s_i|\mathit{salt}_\mathit{index}),\mathit{salt}_\mathit{index}\}$.
Subsequently, the user sends the individual secret $s_i$ to each of the responsible peers to authenticate herself.
Each responsible peer verifies this by hashing the received secret and comparing it to the stored hash.
The message flow of the OTH authentication is shown in Figure~\ref{fig:ac.authentication.OTH}.

It is impossible to determine the secret $s_i$ from the stored hash $h(s_i|\mathit{salt}_\mathit{index})$. Hence, after the initial access and before the first authentication, no subverted peer is able to impersonate the user. However, to authenticate, the user needs to send the secrets $s_i$ to all peers. In a P2P network, the responsible peers for a certain index  might change over time. Therefore, some peers might still be able to collect some or all of the individual secrets. If the attacker manages to collect more than $k$ individual secrets, she can still impersonate the corresponding user.
To prevent this, we introduce our second extension, the so-called \emph{one-time-extension}. With this extension, the individual secrets must only be used once, i.e., we update it with each authenticated operation. More specifically, the user generates a new individual secret for each operation by hashing the current one, i.e., $s'_i= \textrm{HMAC}_s(s_i)$. In an authenticated operation, the user sends $\{s_i,h(s'_i|\mathit{salt}_\mathit{index})\}$ to the $2k+1$ responsible peers. For this to work, the user must additionally store the current individual secrets $s_i$ along with the master secret $s$. The one-time-extension also provides protection against replay attacks as each authenticated message uses new secrets.

\subsection{User Authorization Mechanisms}
\label{sec:ac.authorization}

Any write operation starts with a user executing \emph{put(index, data, auth)} on $2k+1$ peers. The steps for enforcing the access policy during each single write access (put) are as follows: The receiving peer checks whether the requested index is empty, i.e., if there is already an owner or not. If the index is empty, the user of the put operation becomes the owner of this entry and the value is stored under the given index. Specifically, the receiving peer creates a new \emph{dht\_value} data structure and stores the received value in the \emph{data} field. Additionally, the \emph{acl} field is initialized with the minimal ACL, i.e., a single ACL item for the requesting user with the \emph{o} right. In contrast, if there is already an owner for this index, the receiving peer verifies the authenticity of the requesting user with the provided \emph{auth}. Then, the peer checks whether the authenticated user has the right to write to this index. By having this right, the locally stored \emph{data} in the \emph{dht\_value} is overwritten with the received \emph{data}.

A read access is mapped to the execution of \emph{get(index)} on $2k+1$ peers. Each individual operation returns the DHT value for the given \emph{index}. There is no verification of the ACL during a read access (get), i.e., any user can read the value for any index in the DHT. As mentioned before, the read access control is enforced by encrypting the data object with a randomized symmetric data encryption key ($k_d$) and distributing it to the intended users. This can be done out-of-band, e.g., per email or instant messaging. Since in this case the DHT is not involved, we do not need any additional mechanisms to protect the read access. Another possibility to distribute $k_d$ is by using public key cryptography. Here, each user needs a public/private key pair. The key $k_d$ is encrypted with each public key of users who should have read access. As a result, each authorized user gets her individual decryption key. To distribute these individual decryption keys, they are stored in the ACL, more specifically in the key field of the ACL items (cf.\ Listing~\ref{lst:acl_data_structure}) corresponding to the authorized users. When a user performs the get operation for an entry, she gets the DHT value including the corresponding ACL with $k_d$. To decrypt the data object, she first uses her private key to decrypt the encrypted $k_d$ and then uses it to decrypt the data object. Thus, the read access to the data object is enforced by having access to $k_d$ or not. An optimisation for the public key approach would be to include the \emph{auth} parameter with the get operation, i.e., \emph{get(index, auth)}. With this additional information the receiving peer could search for the user's encrypted $k_d$ and reply it together with the data object within the DHT value. Without the \emph{auth}, the user needs to search for her key after receiving the DHT value. Alternatively, the read access can be realized based on Shamir's secret sharing algorithm~\cite{shamir1979share} as done in \cite{6197504}. However, we decided to use the above described method with encryption, as it suits better with our modular approach. If no read access control ist required, we can omit the encryption of the data object altogether (e.g., for DRS in \cite{kikowa2015drs}).

Besides enforcing the write and the read access to data objects, we require a mechanism to administrate the access rights. Thus, we need to provide the owner with a mechanism to delegate the other rights, i.e., read, write and admin. For this, we use \emph{set(index, acl, auth)}. While the put operation modifies the data object, the set operation is intended to modify the ACL. The access control for this operation is also handled on the receiving peer, i.e., with CR, similarly to the put operation. Whenever a peer receives a set operation for a certain \emph{index}, it first verifies whether there is already a value stored for this \emph{index}. If there is no value stored yet, the user who initiated the set operation becomes the owner of this entry. Additionally, any received access rights in \emph{acl} are also applied to this entry. If there is already a value for this index, the peer first verifies whether the requesting user has the appropriate rights to change the ACL of this index. To change read or write access, the requesting user needs at least admin rights, and to change the admin right, she must be the owner.

With the set operation, we can also revoke the access of a user to specific index by removing her from the ACL. Any subsequent write access by the revoked user will be denied by the responsible peers. To revoke a subsequent read access, we need to re-encrypt the data object with a new key which is not known to the revoked user. With our PK approach, this can be achieved directly by using the put and the set operation. Specifically, the user chooses a new encryption key and re-encrypts the data object, which she then stores with a put operation in the DHT. Afterwards, she uses the set operation to set the new rights and the new encrypted data encryption key for each authorized user, i.e., all from before with the exception of the revoked one. Also with the ZKP and the OTH approach, the data object must be re-encrypted and stored in the DHT using a put operation. However, there is no integrated key exchange with the ZKP and OTH approach. Thus, the user must re-distributed this key out-of-band to all remaining authorized users. The costs for revoking a user are the summed effort of a put and a set operation, which we evaluate in the next section.

\subsection{Properties}
\label{sec:ac.properties}

With the above presented approach, we achieve all requirements from Section~\ref{sec:requirements}. By using one of the three proposed authentication mechanisms together with an ACL for each index, we achieve the $A$-requirement. The read protection is achieved by encrypting the data, and the write protection is achieved by replicating the data to $2k+1$ peers and using majority voting. Malicious peers might still write in conflict with the ACL, but their actions remain inert with respect to the entire system.

With our approach, the owner of a DHT entry is uniquely defined and cannot be altered by the attacker ($O$-requirement). To take ownership of an DHT entry, the attacker would need to circumvent the access control or insert herself as the owner during the initial access. However, taking ownership during initial access is secured by sending the request to $2k+1$ peers. If the attacker is not able to subvert a majority of these peers, she cannot modify the initial access with success.

With an individual ACL for each index, we achieve the fine-granularity, i.e., the $G$-requirement. We introduced four different access rights ($o$, $a$, $w$, and $r$) and a new operation -- set -- for delegating the rights to other users. The access to the new set operation is handled with the same mechanisms as with the put or get operation.

To fulfill the $P$-requirement, we use a different authenticator for each index as described above with each authentication mechanism.

The scalability ($S$-requirement) of our approach mainly depends on the total number of messages used. As we show in the next section, our approaches uses a constant factor ($\approx 2k \dots 4k$) of additional messages with respect to a DHT without any access control. As this factor does not depend on the number of peers, our approach scales in the same way as the underlying DHT architecture.

We achieve our $R$-requirement by replicating all DHT values to $2k+1$ different indexes. When reading a DHT entry, we perform a majority voting. Hence, a value can only be modified by altering a majority of all replicas.

\section{Evaluation}
\label{sec:evaluation}

In the following, we discuss the security properties and present an analytical model of our approach. To compare the different authentication mechanisms, we determine the overhead by simulating the involved cryptographic operations.

\subsection{Security Analysis}
\label{sec:evaluation.securityanalysis}
For the security analysis of $k$-rAC, we use the attacker model described in Section~\ref{sec:systemmodel}. As mentioned there, the main goal of the attacker is to circumvent the access control.

In general, for any DHT operation, the user sends a request to $2k+1$ different indexes in the DHT. For this, the user uses the API of $2k+1$ arbitrary but different peers. The $2k+1$ indexes are calculated by a cryptographic hash function and are, therefore, evenly distributed over the entire DHT space. Hence, the probability for housing these $2k+1$ indexes on different peers increases with the total number of peers online. That is, assuming there are enough peers online, each replica is stored on a different peer. Therefore, the $2k+1$ requests will take different paths through the P2P network. In~\cite{heck2016evaluating}, the authors showed that these disjoint paths exist with Kademlia.
To manipulate a single request, the attacker would need to subvert a responsible peer or a peer along the routing path of the request. If the attacker can subvert at most $k$ different peers, she can modify at most $k$ requests (or the corresponding data objects). However, as each DHT operation always operates on $2k+1$ indexes, there are always at least $k+1$ (a majority) of indexes which the attacker cannot modify. Therefore, due to the majority voting, only those values with more than $k$ replications are considered valid rendering the actions of the attacker inert.

Even with more than $k$ subverted peers, our approach gracefully degrades. To manipulate a DHT value, the attacker needs to control the majority of the responsible peers for the corresponding index. However, she cannot freely choose the index in the DHT for which she is responsible, it becomes unlikely that she can subvert exactly the `right' peers for a specific index. Thus, she needs to subvert a much higher number of peers to manipulate at least $k+1$ indexes for a specific index in the DHT.

The PK approach offers an additional security feature, i.e., the signature of the transmitted value. By using a signature, the attacker cannot modify the value in transit through the P2P network. This decreases the attacker's possibilities: to modify a value, she necessarily needs to subvert the responsible peers for a specific index -- subverting peers along the routing path of the request is no longer sufficient. Hence, with the PK approach, we achieve an even better graceful degradation in cases when the attacker can subvert more than $k$ peers.

\subsection{Performance}
\label{sec:evaluation.performance}
Below, we establish an analytical model to determine the performance of our access control scheme with respect to time, message, storage, and computational overhead. We compare our approach, especially our three authentication schemes (AS), with a classical DHT without any access control (DHT w/o AC). To quantify our analytical results, we exemplarily apply well-known and suited cryptographic algorithms to our approach. For this, we
use performance data from the literature and from own measurements.

\subsubsection{Response Time}
\label{sec:evaluation.performance.time_overhead}
We assume that a put or a get operation takes on average $t$ ms for a single operation to succeed in a DHT w/o AC. Our access control scheme does not change this, as we build on top of these operations. Especially, we do not change anything about the underlying mechanisms of the specific DHT implementation, e.g., routing.
In fact, for our $R$-requirement, we must execute at least $2k+1$ classical operations for any authenticated operation. However, these operations are executed in parallel. Hence, the average time to complete an authenticated operation should not vary significantly from the average time $t$ ms of a single operation in a DHT w/o AC. This is only true for the PK and OTH approaches, i.e., $\mathbb{T}_{PK} = \mathbb{T}_{OTH} = t$~ms. With the ZKP approach, we need an additional request/reply. Hence, the expected average time for ZKP is $\mathbb{T}_{ZKP} = 2t$ ms. Two well-known implementations of DHTs are Kademlia and Chord. Kovacevic et al.~\cite{Kovacevic2008} analyzed the average time $t$ for these implementations, yielding $t_{Kademlia} = 100 \ldots 250$ ms and  $t_{Chord} = 450 \ldots 700$ ms. We summarize the results in Table~\ref{tab:evaluation:time}. Accordingly, the PK and the OTH approaches have the same response time as a DHT w/o AC, while the ZKP requires twice as much.

\begin{table}
\center
\begin{tabular}{r|>{\centering}p{3.35cm}|>{\centering\arraybackslash}p{2.3cm}}

  AS  & Response Time $\mathbb{T}_{AS}$ (ms) & Overhead $\Delta \mathbb{T}_{AS}$ \\
  \hline\hline
  DHT w/o AC & $t \approx 100 \ldots 700$ & - \\
  \hline
  PK & $t \approx 100 \ldots 700$ & $0$  \\
  \hline
  ZKP & $2t \approx 200 \ldots 1400$ & $t \approx 100 \ldots 700$ \\
  \hline
  OTH & $t \approx 100 \ldots 700$ & $0$  \\

\end{tabular}
\caption{Response Times}
\label{tab:evaluation:time}
\end{table}

\subsubsection{Message Overhead}
\label{sec:evaluation.performance.message_overhead}
We assume that $y$ messages are required for \emph{put} or \emph{get} in a DHT w/o AC scheme. The PK and OTH approaches use the same number of messages from the requesting peer to a single responsible peer as a put or get operation in a DHT w/o AC. However, since we need to send the request to $2k+1$ different responsible peers, we require in total $m_{PK}=m_{OTH}=(2k+1)\cdot y$ messages. Thus, the overhead of these approaches can be estimated with $\Delta m_{PK}=\Delta m_{OTH} = m_{PK} - y = 2ky$. The ZKP approach requires an additional request/reply pair of messages. Hence, the total number of messages is $m_{ZKP}=(2k+1)\cdot 2y$, and, therefore, the overhead is $\Delta m_{ZKP} = m_{ZKP} - y = 4ky + y$. These results are shown in Table~\ref{tab:evaluation:message_overhead}. Comparing our approaches, PK and OTH use the same number of messages, while ZKP uses roughly twice as much.

The next metric is the message size overhead. In our access control scheme, this is the additional \emph{auth}-parameter in each operation. In the PK approach, it contains a public key, a counter, and a digital signature, i.e., $\mathit{auth}=\{\mathit{pk}, \mathit{ctr}, \mathit{signature}\}$ (cf.~Section~\ref{sec:ac.authentication}). We use a 32 bits counter for preventing replay attacks. The sizes of the public key and the signature depend upon the used asymmetric cryptographic algorithm. Exemplarily, we use RSA~2048 and ECC (Elliptic Curve Cryptography)~224, which provide roughly the same security. Here, the key lengths refer to the bit lengths of the moduli (RSA) and the subgroup (ECC). However, the actual number of bits required to store a public key for both algorithms is larger, as the public key contains additional parameters~\cite{lenstra2001selecting}. To determine the size of the \emph{auth}-parameter, we used OpenSSL~\cite{viega2002network} to generate a new key pair for both cryptosystems and stored the public key in the binary DER format~\cite{x690}. This yields 294 bytes for the RSA key and 80 bytes for the ECC key. The signature in the authenticator is an encrypted hash of the data object concatenated with the counter, i.e., $\mathit{signature} := \mathrm{encrypt}_{sk}(h(\mathit{data}|\mathit{ctr}))$. The size of the signature also depends on the specific hash function used. Here, we used SHA-256 (32 bytes) exemplarily and encrypted the resulting hash with both asymmetric algorithms. This results in 256 bytes for RSA and 63 bytes for the ECC signature. Hence, our total overhead is $\Delta s_{RSA} = 4 + 294 + 256 = 554$ bytes and $\Delta s_{ECC}= 4 + 80 + 63 = 147$ bytes. With the ZKP approach, we only consider the final message when calculating the size overhead. The first and the second message are new messages and already included in $\Delta m_{ZKP}$. Additionally, these messages are small: the first message contains the initialization of the protocol and the second message contains the challenges from the responsible peer, which can be represented as $n$ bits. A sound value for $n$ is, e.g., $20$. In comparison, a put or a get message contains at least the DHT index, e.g., 160 bit. The final message contains $n$ responses to the challenges from the second message (cf. Figure~\ref{fig:ac.authentication.ZKP}). A single response represents a large number, big enough to make the calculation for the discrete logarithm infeasible. Hence, around 200 decimal places or 665 bits $\approx$ 83 bytes. This yields an overhead of $n\cdot 83$ bytes, i.e., with $n=20$, the overhead is 1660 bytes. Finally, with the OTH approach, each message contains two additional hashes. By using again SHA-256 as the hash function, we get an overhead of $2\cdot 32 = 64$ bytes. We summarize the results of this metric in the second column of Table~\ref{tab:evaluation:message_overhead}. Comparing our approaches, OTH and PK with ECC use the smallest overhead with respect to message size.

\begin{table}
\center
\begin{tabular}{r|>{\centering}p{3.10cm}|>{\centering\arraybackslash}p{3.10cm}}

  AS  & Number $\Delta m_{AS}$ & Size $\Delta s_{AS}$ (B)  \\
  \hline\hline
  DHT w/o AC & $y$ & -  \\
  \hline
  PK with RSA & $2ky$ & $554$  \\
  \hline
  PK with ECC & $2ky$ & $147$  \\
  \hline
  ZKP & $(4k+1)\cdot y$ & $n \cdot 83$ \\
  \hline
  OTH & $2ky$ & 64  \\

\end{tabular}
\caption{Message Overhead}
\label{tab:evaluation:message_overhead}
\end{table}

\subsubsection{Storage Overhead}
\label{sec:evaluation.performance.storage_overhead}

With respect to a DHT w/o AC, the storage overhead differs for the responsible peers and users. We assume that the used DHT uses a 160-bit hash function (e.g., SHA-1). Hence, for storing the index to a specific value in the DHT w/o AC, the storage requirement for any user is 20 bytes. Furthermore, we assume that each responsible peer uses a data structure (e.g., a hash table) to locally store the DHT values.

\textbf{User's Storage Overhead} Using the PK approach, the user needs to store a different public/private key pair for each index she has access to. Additionally, she must store a counter for the freshness of the messages. We assume a 32-bit counter, hence 4 bytes. Since each index is mapped for the $k$-resilience to $2k+1$ different positions in the DHT, she must store those positions, i.e., $(2k+1) \cdot 20$ bytes. Similarly to our analysis of the message size overhead, we used OpenSSL to determine the size of a RSA and a ECC key pair in DER format, i.e., 1192 bytes for RSA and 113 bytes for ECC. Hence, the total amount the user needs to store is $\mathbb{S}_{RSA} = 1192 + 4 +20 \cdot (2k+1) = 1216 + 40k$ or  $\mathbb{S}^U_{ECC} = 113 + 4 +20 \cdot (2k+1) = 137 + 40k$. Since the user would need to store one index (20 bytes) also in a DHT w/o AC, the overhead for the PK approach is $\Delta \mathbb{S}^U_{RSA} = 1196 + 40k$ or  $\Delta \mathbb{S}^U_{ECC} =  117 + 40k$.

With the ZKP approach, the user stores only her individual secret for each DHT entry she has access to. As described in our analysis of the message size overhead, we require  $\approx 83$ bytes for the responses. These responses are either the square of the secret or a product of two large numbers. Thus, to store the secret, we require $\approx 45$ bytes. Hence, the total amount she stores is $\mathbb{S}^U_{ZKP} = 45 + 20 \cdot (2k+1)$. Similarly as before, the overhead is $20$ bytes less, since one index would be stored anyway. Thus, the overhead of the ZKP approach is $\Delta \mathbb{S}^U_{ZKP} = 25 + 40k$.

With OTH, the user stores once the master key (we assume 32 byte). For each DHT entry she has access to, she stores a salt (16 byte), $2k+1$ current secrets (each 32 bytes), and the corresponding $2k+1$ positions (cf.\ OTH in Section \ref{sec:ac.authentication}). Hence, the total storage amount for a single index is $\mathbb{S}^U_{OTH} = 16 + (2k+1)\cdot 32 + (2k+1)\cdot 20$. Hence, the overhead for each index is $\Delta \mathbb{S}^U_{OTH} = 48 + 104k$ bytes.

\begin{figure}
  \centering
  \includegraphics[width=1\columnwidth]{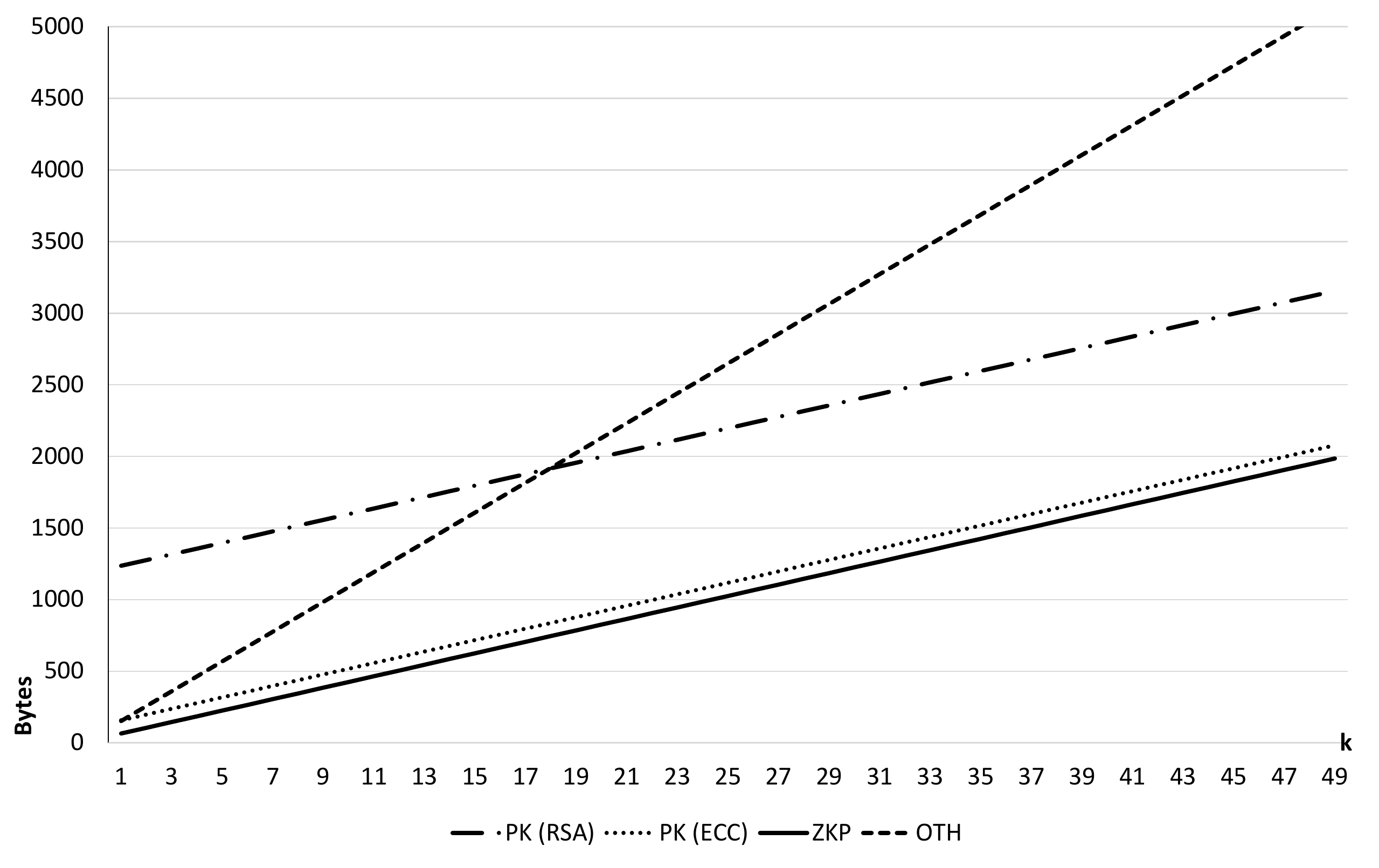}\\
  \caption{Comparison of User's Storage Overhead $\Delta \mathbb{S}^U_{AS}$}\label{fig:evaluation.performance.storage_overhead_user}
\end{figure}

In Table \ref{tab:evaluation:Storage_overhead}, we summarize the storage overhead for the user having access to $i$ different DHT entries.
As with any ACL mechanism, the storage overhead for the user depends linearly from the number of DHT entries she has access to. However, with $k$-rAC, it also depends on the resilience value $k$ as shown in Figure \ref{fig:evaluation.performance.storage_overhead_user}. Accordingly, the ZKP approach has the least storage requirement. The PK approach with ECC uses only slightly more storage and is also well suited, even for higher $k$ values. Since ECC includes the key exchange, it offers more functionality than the ZKP approach. On the other hand, the PK approach with RSA should be avoided due to the overall higher storage requirement. In contrast to the other approaches, the OTH approach does not scale well with higher $k$ values (higher slope). For $k>17$, it even requires more storage than the PK approach with RSA.

\textbf{Peer's Storage Overhead} With the PK approach, the responsible peer stores additional meta information for each authorized user in the ACL according to Listing \ref{lst:acl_data_structure}:  the encrypted symmetric key, the user rights, and the authenticator. We assume a symmetric key of 128 bits (16 bytes) and one byte for the user rights. Encrypting the symmetric key yields 256 bytes for RSA and 64 bytes for ECC. The size of the authenticator depends on the used authentication mechanism. With the PK approach, the authenticator comprises the public key and a sliding window of accepted counters. We used 32 bits for the counter and a window of 32 entries, i.e., $4 \cdot 32 = 128$ bytes. As before, the size of the public key is 294 bytes for RSA and 80 bytes for ECC. Hence, the storage overhead for each ACL item is $\Delta \mathbb{S}^P_{RSA} = 256 + 1 + 128 + 294 = 679$ bytes for RSA and $\Delta \mathbb{S}^P_{ECC} = 64 + 1 + 128 + 80 = 273$ bytes for ECC.

With the ZKP and the OTH approach, the peers do not store a symmetric encryption key, since they exchange it out-of-band. Hence, with the ZKP approach the peers stores only the square of the secret and the user rights, i.e., $\Delta \mathbb{S}^P_{ZKP} = 83 +1 = 84$ bytes for each ACL item.

Similarly, with the OTH approach, the peer only stores a hash value (23 bytes), a salt value (16 bytes) and the user rights (1 byte). Thus, the storage overhead is $\Delta \mathbb{S}^U_{OTH} = 32 +16 + 1 = 49$ bytes. In Table \ref{tab:evaluation:Storage_overhead}, we summarize the storage overhead for a peer with an ACL containing $u$ items. The storage overhead for the peers is independent of $k$ and scales linearly with the number of ACL items. We get the lowest overhead with the OTH and the ZKP approach, while the PK approaches use three to fourteen times more storage.

\begin{table}
\center
\begin{tabular}{r|>{\centering}p{3.10cm}|>{\centering\arraybackslash}p{3.10cm}}

  AS  & $\Delta \mathbb{S}^U_{AS}$ (B) & $\Delta \mathbb{S}^P_{AS}$ (B)    \\
  \hline\hline
  PK with RSA & $i \cdot (1196 + 40k)$ & $u \cdot 679$ \\
  \hline
  PK with ECC & $i \cdot (117 + 40k)$ & $u \cdot 273$ \\
  \hline
  ZKP  &  $i \cdot (25 + 40k)$ & $u \cdot 84$ \\
  \hline
  OTH  & $32 + i \cdot (48 + 104k)$ & $u \cdot 49$ \\

\end{tabular}
\caption{Storage Overhead}
\label{tab:evaluation:Storage_overhead}
\end{table}

\subsubsection{Computational Overhead}
\label{sec:evaluation.performance.computational_overhead}
To evaluate the computational overhead, we differentiate three metrics: First, we analyze the initial effort for the user on first access to any index in the DHT, i.e., $\mathbb{U}^0_{AS}$. Secondly, we determine the effort for the user during any subsequent authenticated access, i.e., $\mathbb{U}^i_{AS}$. Finally, we analyze the computational overhead of the responsible peers, i.e., $\mathbb{P}_{AS}$. For each of these metrics, we consider the three authenticated operations (put, get, and set) separately as they cause different effort.

\textbf{User's Initial Effort} Initially, every DHT entry is empty and does not have an owner. Albeit a get operation for an empty entry is not forbidden, it returns a \emph{null} data object. Hence, in our system an initial get operation causes no overhead with respect to a DHT w/o AC.

For a put operation using the PK approach, the user initially creates a key pair for the accessed index. This key generation operation ($\mathit{KG}$) depends on the used asymmetric cryptographic algorithm, e.g., with RSA, it involves finding two large primes. For the initial access, we do not sign the data object, as there is no way for a peer to securely verify this signature (the corresponding public key is in the same message). We achieve the security of the initial put operation with the $k$-resilience property of our approach. For the read access, the data object is encrypted with a symmetric encryption algorithm ($SO$). For $a$ different ACL items, we encrypt this symmetric encryption key with the public key of each authorized user ($\mathit{AO}_{pk}$). At the very least, the corresponding encryption key is encrypted with the owner's public key. Due to our clear separation of managing data and managing the ACL, this prepared ACL must be sent with a subsequent set operation.
Hence, the initial complexity of a put operation for the user is $\mathbb{U}^0_{PK} = \mathit{KG} + \mathit{SO} + a \cdot \mathit{AO}_{pk}$. Similarly, for an initial set operation, we generate a new key pair, i.e., $\mathit{KG}$. However, as there is no data object in a set operation, there is also no $\mathit{SO}$ involved. Nevertheless, we still need to generate a symmetric encryption key for encrypting the data object later on. Hence, the user's initial complexity for a set operation is $\mathbb{U}^0_{PK} = \mathit{KG}  +  a \cdot \mathit{AO}_{pk}$.

For an initial put or set operation using the ZKP approach, the user generates a random number and calculates its modular square. Assuming that the effort for generating a random number is negligible, we only consider the modular operations ($\mathit{MO}$). For a put operation, the data object is also encrypted with an $\mathit{SO}$ operation. Hence, the user's initial effort for put is $\mathbb{U}^0_{ZKP} = \mathit{MO} + \mathit{SO}$ and for set $\mathbb{U}^0_{ZKP} = \mathit{MO}$.

With the OTH approach, the user initially calculates $2k+1$ individual hashes for put or set. Similarly as before, the data object is encrypted for put with an $\mathit{SO}$ operation. In summary, the user's initial effort for the put operation is $\mathbb{U}^0_{OTH} = (2k+1)\cdot\mathit{HO} + \mathit{SO}$ and for the set operation is $\mathbb{U}^0_{OTH} = (2k+1)\cdot\mathit{HO}$.

We summarize the results for the user's initial effort for all cases in Table~\ref{tab:evaluation:initial_user_computational_overhead}.

\begin{table}
\center
\begin{tabular}{r|>{\centering}p{3.40cm}|>{\centering\arraybackslash}p{3.40cm}}

  \multirow{2}{*}{AS}  & \multicolumn{2}{c}{User $\mathbb{U}^0_{AS}$}  \\
   & put & set  \\
  \hline\hline
  PK & $\mathit{KG} + \mathit{SO} + a \cdot \mathit{AO}_{pk}$ & $\mathit{KG} + a \cdot \mathit{AO}_{pk}$  \\
  \hline
  ZKP & $\mathit{MO} + \mathit{SO}$ & $\mathit{MO}$  \\
  \hline
  OTH &  $(2k+1)\cdot\mathit{HO} + \mathit{SO}$ & $(2k+1)\cdot\mathit{HO}$ \\

\end{tabular}
\caption{Initial User Computational Overhead}
\label{tab:evaluation:initial_user_computational_overhead}
\end{table}

\textbf{User's Subsequent Effort} For any subsequent get operation, the user does not have to authenticate, since the read access control is done with CQ. She only has to retrieve the data object and decrypt it. Hence, the user's subsequent effort for a get operation includes for all three authentication schemes an $\mathit{SO}$ operation for decrypting the data object. With the ZKP and the OTH approach, the decryption key is distributed out-of-band. Thus, there is no additional effort in these cases. However, for the PK approach, we first need to decrypt the encryption key using an asymmetric cryptographic operation with the user's private key ($\mathit{AO_{sk}}$). Hence, the user's subsequent effort can be determined with $\mathbb{U}^i_{PK} = \mathit{AO}_{sk} + \mathit{SO}$ for the PK approach and $\mathbb{U}^i_{ZKP} = \mathbb{U}^i_{OTH} = \mathit{SO}$ for the ZKP and the OTH approach.

For a put operation with the PK approach, we encrypt the data object with an $\mathit{SO}$ operation. After that, the encryption key is encrypted with the public keys of all authorized users, i.e., $a \cdot \mathit{AO}_{pk}$. Again, this prepared ACL must be sent with a subsequent set operation. Here, each message also contains a signature which requires hashing ($\mathit{HO}$) and an asymmetric operation with the private key ($\mathit{AO}_{sk}$). Hence, the user's total subsequent effort can be determined by $\mathbb{U}^i_{PK} = \mathit{SO} + a \cdot \mathit{AO}_{pk} + \mathit{HO} + \mathit{AO}_{sk}$. For the set operation with the PK approach, the user's subsequent effort is similar to the put operation but without the need to encrypt a data object, i.e., without the $\mathit{SO}$.

With the ZKP approach, we also encrypt the data object for a put operation. To authenticate the operation, the user calculates $n$ responses to the challenges from each of the $2k+1$ peers. For this, she first chooses a random number $r$ and calculates its modular square ($n$ MO). Depending on the peer's challenge (i.e., 1 or 0), the user replies with the modular product of her secret with $r$ or only $r$. On average, there will be $n/2$ 1s in the challenge yielding additional $n/2$ MOs. This involves on average $n \cdot 1.5 \cdot (2k+1)$ modular operations ($\mathit{MO}$). Thus, the users's subsequent effort for a put operation is $\mathbb{U}^i_{ZKP} = \mathit{SO} + (2k+1)\cdot n \cdot \mathit{MO}$. Again, the set operation needs the same effort but without the encryption of the data object, i.e., $\mathbb{U}^i_{ZKP} = (2k+1)\cdot n \cdot 1.5 \cdot \mathit{MO}$.

Finally, with the OTH approach, a put operation also requires encrypting the data object. For the authentication, we need to calculate $2k+1$ individual hashes. Hence, the user's subsequent effort for a put operation is $\mathbb{U}^i_{OTH} = \mathit{SO} + (2k+1)\cdot\mathit{HO}$. As with the other two approaches, the only difference for the set operation is the lack of the data object encryption, i.e., $\mathbb{U}^i_{OTH} = (2k+1)\cdot\mathit{HO}$.

In all three approaches, the user performs a majority voting over all received results. However, this operation is negligible in comparison to computational complex operations like modular arithmetic or hash calculations. We summarize our analytical results of the user's subsequent effort in Table~\ref{tab:evaluation:subsequent_user_computational_overhead}.

\begin{table*}
\center
\begin{tabular}{r|>{\centering}p{5.05cm}|>{\centering}p{5.05cm}|>{\centering\arraybackslash}p{5.05cm}}

  \multirow{2}{*}{AS}  & \multicolumn{3}{c}{User $\mathbb{U}^i_{s}$}  \\
   & get & put & set  \\
  \hline\hline
  PK & $\mathit{AO}_{sk} + \mathit{SO}$ & $\mathit{SO} + a \cdot \mathit{AO}_{pk} + \mathit{HO} + \mathit{AO}_{sk}$ & $a \cdot \mathit{AO}_{pk} + \mathit{HO} + \mathit{AO}_{sk}$  \\
  \hline
  ZKP & $\mathit{SO}$ & $\mathit{SO} + (2k+1)\cdot n \cdot 1.5 \cdot \mathit{MO}$ & $(2k+1)\cdot n \cdot 1.5 \cdot \mathit{MO}$ \\
  \hline
  OTH & $\mathit{SO}$ & $\mathit{SO} + (2k+1)\cdot\mathit{HO}$ & $(2k+1)\cdot\mathit{HO}$  \\

\end{tabular}
\caption{Subsequent User Computational Overhead}
\label{tab:evaluation:subsequent_user_computational_overhead}
\end{table*}

\textbf{Peer's Effort} Our final metric is the computational overhead for the responsible peers. As we use CQ for the get operation (cf.~Section~\ref{sec:ac.authentication}), we do not need to authenticate the get requests. Hence, there is no overhead for the responsible peers in comparison to a DHT w/o AC.

For a put and a set operation, the responsible peer needs to authenticate the user. For the peer, there is no difference whether it is an put or a set operation. With the PK approach, the peer must verify the signature to authenticate the request. This involves decrypting the signature with the public key and hashing the data object. Hence, the resulting complexity is $\mathbb{P}_{PK} = \mathit{HO} + \mathit{AO}_{pk}$. With the ZKP and the OTH approach, the integrity of the data object is not verified (we rely on the $k$-resilience). Hence, there is no hash operation involved. The authentication with the ZKP approach involves calculating $n$ times the modular square of the received response $y$ ($n$ MOs) and comparing it either with the modular product $v \cdot r^2$ or only with $r^2$, i.e., $y^2 \equiv v^c \cdot r^2$ (mod $p$) for $c \in \{0,1\}$ (cf.~Figure~\ref{fig:ac.authentication.ZKP}). Whether we need to perform the modular product depends on the challenge $c$. Hence, assuming that on average there are about half of the challenge bits 1s, we need to perform $n/2$ MOs. This yields an average complexity of $\mathbb{P}_{ZKP} = n \cdot 1.5 \cdot \mathit{MO}$. Finally, with the OTH approach, a responsible peer only performs one hash calculation to verify the authenticity of the user, i.e., $\mathbb{P}_{OTH} = \mathit{HO}$. We summarize our results in Table~\ref{tab:evaluation:peer_computational_overhead}.

\begin{table}
\center
\begin{tabular}{r|>{\centering}p{2.12cm}|>{\centering}p{2.12cm}|>{\centering\arraybackslash}p{2.12cm}}

  \multirow{2}{*}{AS}  & \multicolumn{3}{c}{Peers $\mathbb{P}_{s}$}  \\
   & get & put & set  \\
  \hline\hline
  PK & 0 & $\mathit{AO}_{pk}  + \mathit{HO}$ & $\mathit{AO}_{pk}  + \mathit{HO}$  \\
  \hline
  ZKP & 0 & $n \cdot 1.5 \cdot \mathit{MO}$ & $n \cdot 1.5 \cdot \mathit{MO}$ \\
  \hline
  OTH & 0 & $\mathit{HO}$ & $\mathit{HO}$  \\

\end{tabular}
\caption{Peer Computational Overhead}
\label{tab:evaluation:peer_computational_overhead}
\end{table}

\subsubsection{Simulation}
\label{sec:evaluation.performance.simulation}
To quantify the results of our analytical analysis and to compare the three proposed authentication mechanisms, we implemented the cryptographic operations by using the built-in cryptographic engines from the cryptography architecture~\cite{javaCryptoDocu} of Java SE 7. We measured the time for each operations by taking the average of one Million executions of the operation on a workstation (Intel i7-4900 MQ, 2.8 GHz, 32 GB RAM). Although the absolute values will vary on different computer systems, the measurements allow to compare the proposed authentication mechanisms. For the ECC-measurements, we used the elliptic curve integrated encryption system with AES ("ECIESwithAES"), which uses ECC together with AES to build an asymmetric cryptosystem. In cases where only a signature is required, it would suffice to use the Elliptic Curve Digital Signature Algorithm (ECDSA). However, we evaluated this and found that ECDSA is slightly slower than ECIESwithAES. Therefore, we did not follow this further. The time for SO and HO depends on the number of bytes encrypted or hashed. We used 32 kB as an average size. We assume that data objects stored in a DHT are rather small and usually fit into a UDP packet (max 64 kB). However, even with bigger data objects, the time needed to perform SO or HO is small in comparison to other operations. This is probably caused by the fact that the used processor uses hardware acceleration for these operations (AES-NI). Nevertheless, we acknowledge that for big data objects (several megabytes to gigabytes) this operation can become a significant factor~\cite{garrison2016practicality}. For MO, we used a big integer number with 200 decimal places, i.e., 665 bits. With this size, the calculation of the discrete logarithm is impractical. In Table~\ref{tab:evaluation:time_simulation}, we summarize the measured times in microseconds.

There are three parameters in our system, namely $n$, $k$, and $a$. The parameter $n$ is the number of challenges with our ZKP approach. With $n=20$, there is  only one in a Million ($1:2^{20}$) chance of fooling the AS. The parameter $k$ is the number of tolerated subverted peers. This parameter highly depends on the possibilities of the attacker to add peers under his control to the network. For our evaluation, we choose exemplarily $k=20$. Finally, the parameter $a$ is the amount of authenticated users in an ACL. Exemplarily, we used $a=10$.

\begin{table}
\center
\begin{tabular}{r|>{\centering\arraybackslash}p{5.40cm}}

   Operation & Time \\
  \hline\hline
  $\mathit{KG}$ (RSA 2048) & $317092\ \mu s$ \\
  \hline
  $\mathit{KG}$ (ECC 224) & $685\ \mu s$  \\
  \hline
  $\mathit{AO}_{sk}$ (RSA 2048) & $5135\ \mu s$ \\
  \hline
  $\mathit{AO}_{sk}$ (ECC 224) & $170\ \mu s$ \\
  \hline
  $\mathit{AO}_{pk}$ (RSA 2048) & $148\ \mu s$ \\
  \hline
  $\mathit{AO}_{pk}$ (ECC 224) & $380\ \mu s$ \\
  \hline
  $\mathit{HO}$ (SHA 256) & $176\ \mu s$ (32 kB)\\
\hline
  $\mathit{SO}$ (AES 128) & $33\ \mu s$  (32 kB)\\
\hline
  $\mathit{MO}$ (665 Bits) & $4\ \mu s$\\
\end{tabular}
\caption{Measurement Results}
\label{tab:evaluation:time_simulation}
\end{table}

\begin{figure}
  \centering
  \includegraphics[width=1\columnwidth]{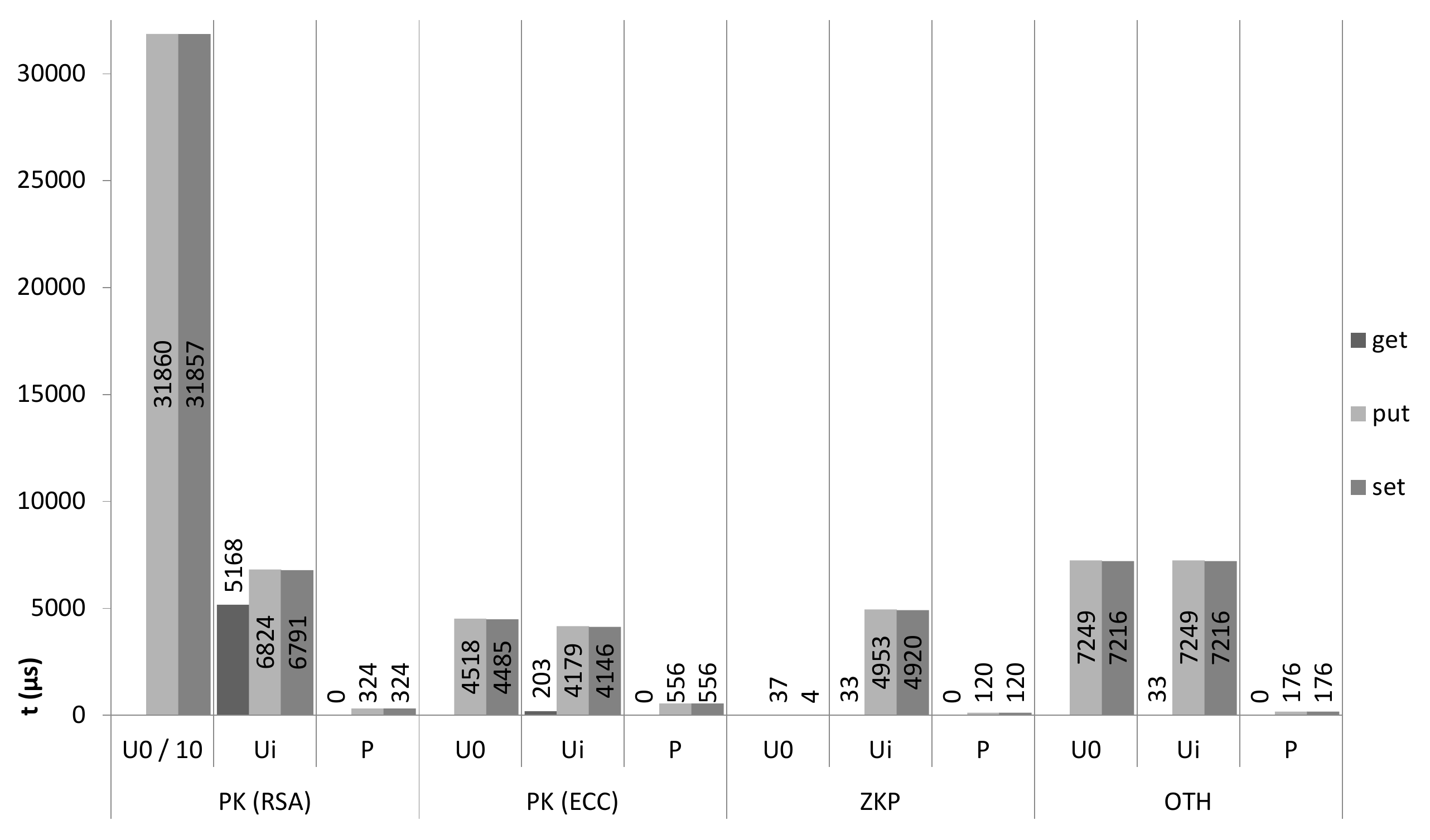}\\
  \caption{Comparison of Authentication Mechanisms}\label{fig:evaluation.performance.simulation.auth}
\end{figure}

In Figure~\ref{fig:evaluation.performance.simulation.auth}, we present the resulting computational overhead of our authentication mechanisms. For this, we used the measured times from Table~\ref{tab:evaluation:time_simulation} and the exemplary values for the parameters from above together with the analytical model. According to these results, the overhead for the peer varies from $100\ \mu s$ to approximately $600\ \mu s$. This overhead is negligible in comparison to approximately $200\ ms$ roundtrip times for DHT operations~\cite{Kovacevic2008}.

When comparing the computational overhead of the get operation, we must consider that the PK approaches include the key distribution. Thus, a higher overhead for the PK approaches is acceptable as they offer additional features. Nevertheless, we see in Figure~\ref{fig:evaluation.performance.simulation.auth} that the overhead for the get operation is negligible for all authentication mechanisms. The highest overhead for the get operation is from the PK approach with RSA, where a get operation needs approximately $5\ ms$. Although this is still acceptable, the PK approach with RSA additionally generates a notably higher overhead for \emph{put} or \emph{set} during the initial access. Therefore, we suggest to use ECC instead of RSA and consider in the remainder of this analysis only the PK approach based on ECC.

For put and set, the OTH approach generates the highest overhead -- both on initial access and on subsequent accesses. This is an interesting result, as the overhead for this authentication mechanism is similar to the overhead of hash chains. Hash chains are usually preferred over asymmetric cryptography due to their lower computational overhead. However, with OTH, we need $2k+1$ individual secrets to tolerate $k$ subverted peers. Therefore, its advantage disappears for a certain $k$. By ignoring the overhead for the key distribution using the PK with ECC approach (by setting $a=0$), PK outperforms OTH already for $k \approx 0.5$. Also for the ZKP approach, we determined that for a $k>1$ the PK approach with ECC outperforms ZKP for any subsequent access. However, the major advantage of ZKP is the low overhead during initial access, where it outperforms all other approaches. Additionally, on subsequent accesses, ZKP needs approximately $1.5$ times less resources than OTH. However, this is just the computational overhead -- ZKP requires an additional message pair (cf.~\ref{sec:evaluation.performance.message_overhead}).

In summary, the PK approach with ECC is a general approach offering the most features with acceptable overhead. For scenarios with more initial than subsequent accesses and without controlled read access, ZKP is a resource friendly alternative. An example for such a scenario is the DRS from \cite{kikowa2015drs}.

\section{Conclusion}
\label{sec:conclusion}

In this paper, we presented $k$-rAC, a novel fine-grained $k$-resilient access control for DHTs. Currently existing DHTs cannot be used in privacy aware applications because of the lack of a reliable and practical access control. With our access control, we open the door for using a DHT in such applications. In our approach, the read or write access to each index in the DHT can be regulated individually. We determined the owner of a DHT entry by the initial access and provided privacy-aware mechanisms to delegate access rights to other users. The security of our approach is based on $k$-resilience, i.e., the access control cannot be circumvented as long as the attacker is not able to subvert more than $k$ peers. Additionally, even in cases when the attacker is able to subvert more than $k$ peers, our approach gracefully degrades. This is caused by the fact that the attacker needs to subvert specific peers in order to circumvent the access control for a specific index.

In our evaluation, we compared three different authentication mechanisms in terms of response time, message, storage and computational overhead. We showed that the caused overhead for all three approaches is acceptable. Since the approaches have different advantages and disadvantages, it depends on the requirements of a specific scenario which of the authentication mechanisms should be used. The suitable authentication mechanism can be determined with the presented analytical model.



In the future, we will extend our fine-grained access control scheme to additionally support permissions for arbitrary groups and everyone similar to the flexible access control known from UNIX for files. For this, we plan to store the mapping of users to groups in the DHT.





%

\bibliographystyle{plain}
\bibliography{Bibliography}

\end{document}